\documentclass[12pt]{article}
\usepackage{epsf}
\usepackage{epsf}
\usepackage{amsmath}
\usepackage{amsfonts}
\newcommand{\hx}{\widehat{x}}

\newcommand{\hv}{\widehat{v}}

\begin{document}
\setlength{\parskip}{2ex} \setlength{\parindent}{0em}
\setlength{\baselineskip}{3ex}
\newcommand{\onefigure}[2]{\begin{figure}[htbp]
         \caption{\small #2\label{#1}(#1)}
         \end{figure}}
\newcommand{\onefigurenocap}[1]{\begin{figure}[h]
         \begin{center}\leavevmode\epsfbox{#1.eps}\end{center}
         \end{figure}}
\renewcommand{\onefigure}[2]{\begin{figure}[htbp]
         \begin{center}\leavevmode\epsfbox{#1.eps}\end{center}
         \caption{\small #2\label{#1}}
         \end{figure}}
\newcommand{\comment}[1]{}
\newcommand{\myref}[1]{(\ref{#1})}
\newcommand{\secref}[1]{sec.~\protect\ref{#1}}
\newcommand{\figref}[1]{Fig.~\protect\ref{#1}}
\newcommand{\mathbold}[1]{\mbox{\boldmath $\bf#1$}}
\newcommand{\mJ}{\mathbold{J}}
\newcommand{\momega}{\mathbold{\omega}}
\newcommand{\bz}{{\bf z}}
\def\bbbz{{\sf Z\!\!\!Z}}
\newcommand{\PP}{\mbox{I}\!\mbox{P}}
\newcommand{\FF}{I\!\!F}
\newcommand{\bbbc}{\mbox{C}\!\!\!\mbox{I}}
\def\sl2z{SL(2,\bbbz)}
\newcommand{\bbbq}{I\!\!Q}
\newcommand{\be}{\begin{equation}}
\newcommand{\ee}{\end{equation}}
\newcommand{\bea}{\begin{eqnarray}}
\newcommand{\eea}{\end{eqnarray}}
\newcommand{\nn}{\nonumber}
\newcommand{\unit}{1\!\!1}
\newcommand{\half}{\frac{1}{2}}
\newcommand{\shalf}{\mbox{$\half$}}
\newcommand{\transform}[1]{
   \stackrel{#1}{-\hspace{-1.2ex}-\hspace{-1.2ex}\longrightarrow}}
\newcommand{\inter}[2]{\null^{\#}(#1\cdot#2)}
\newcommand{\lprod}[2]{\vec{#1}\cdot\vec{#2}}
\newcommand{\mult}[1]{{\cal N}(#1)}
\newcommand{\Bn}{{\cal B}_N}
\newcommand{\B}{{\cal B}}
\newcommand{\Beight}{{\cal B}_8}
\newcommand{\Bnine}{{\cal B}_9}
\newcommand{\Eman}{\widehat{\cal E}_N}
\newcommand{\C}{{\cal C}}
\newcommand{\Q}{Q\!\!\!Q}
\newcommand{\comp}{C\!\!\!C}

\noindent

\thispagestyle{empty} {\flushright{\small
UTTG-15-01\\hep-th/0109214\\}}

\vspace{.3in}
\begin{center}\Large {\bf Discrete Symmetries of the Superpotential and 
Calculation of Disk Invariants}
\end{center}

\vspace{.1in}
\begin{center}
Amer Iqbal\,\,\,  and \,\,\,Amir-Kian Kashani-Poor 
\vspace{.05in}

Theory Group, Department of Physics,\\
University of Texas at Austin,\\
Austin, TX, 78712. \vspace{.2in}

\end{center}

\vspace{0.1in}

\begin{abstract}
The integrality of Ooguri-Vafa disk invariants is verified using
discrete symmetries of the superpotential of the mirror
Landau-Ginzburg theory to calculate quantum corrections to the
boundary variables. We show that these quantum corrections are
completely determined if we assume that the discrete symmetry of the
superpotential also holds in terms of the quantum corrected
variables. We discuss the case of local $\mathbb{P}^{2}$ blown up at
three points and local $\mathbb{F}_{2}$ blown up at two points in
detail.
\end{abstract}

\newpage

\thispagestyle{empty}

{\scriptsize

\tableofcontents

}

\pagenumbering{arabic}

\section{Introduction}

The calculation of topological string amplitudes with boundaries has
received a lot of attention recently\cite{AV,AKV,mayr,GJS}. Besides
being of mathematical interest, these amplitudes capture exactly
certain terms in the effective ${\cal N}=1$ four dimensional theory
\cite{BCOV}. The calculation of disk amplitudes has been carried out
both using mirror symmetry \cite{AV,AKV} and directly using
localization techniques similar to the ones used for calculating
closed string Gromov-Witten invariants \cite{KL,GZ,LS}. These
calculations have verified the integrality of open string invariants
defined in \cite{OV}. Also, large N duality with Chern-Simons theory 
\cite{CS}
has lead to the verification, in certain cases, of the general
structure of amplitudes as predicted in \cite{MLV,OV}, for all genera
\cite{MV}. More recently it has been proposed, and verified in certain
cases, that disk amplitudes can be obtained from genus zero closed
topological string amplitudes of a Calabi-Yau fourfold \cite{mayr}.

The most extensive check of the integrality of open string invariants
was carried out in \cite{AKV}. \figref{TToric}(a) shows the toric
data of the geometries discussed in that paper. As in the closed string
case, it is necessary to express the disk amplitude in the ``flat
coordinates'' of the boundary theory in order to obtain integer valued
invariants.  The flat coordinates of the boundary theory are related
to the classical area of the disk through closed string quantum
corrections. It was shown in \cite{AKV} that these quantum corrections
to the classical area of the disk can be expressed as integrals over
certain cycles on the Riemann surface which parameterizes the position
of the mirror D-brane.

In this paper, we show that quantum corrections to the boundary
variables can be calculated using discrete symmetries of the
superpotential of the Landau-Ginzburg theory which is mirror to the
Calabi-Yau threefold.  \figref{TToric}(c) gives the toric data of the
geometries we will discuss in this paper, local $\mathbb{P}^{2}$ blown up at three points
and local $\mathbb{F}_{2}$ blown up at two points (both with four
K\"ahler parameters). Taking suitable
limits of the K\"ahler parameters, we can blow down exceptional curves
and therefore obtain results for the blown down geometries depicted in a) and b) with no further effort. We will show that quantum corrections obtained
using the discrete symmetries lead to integer invariants.

\onefigure{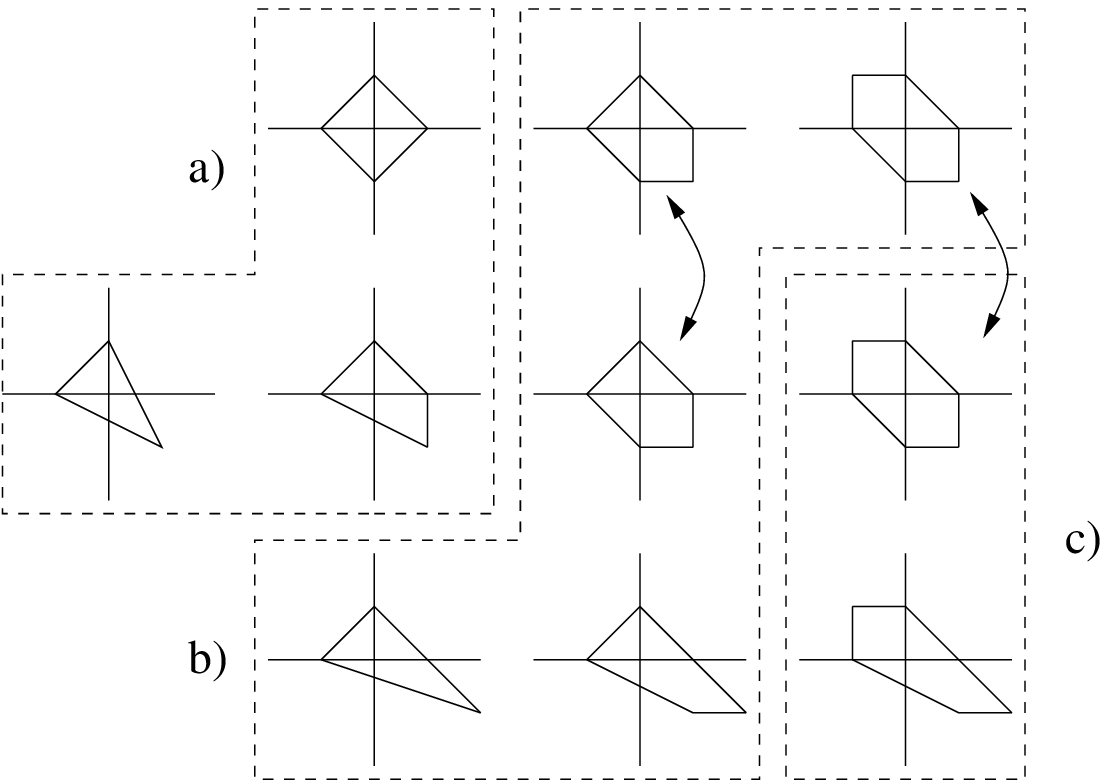}{a) Toric data for $\mathbb{F}_0$ ($\mathbb{P}_1 \times \mathbb{P}_1$), $\mathbb{P}^{2}$, and $\mathbb{F}_1$ ($\mathbb{P}^{2}$ blown up at one point), b) and c) blowups of $\mathbb{F}_0$ and of $\mathbb{F}_1$, $\mathbb{F}_2$ and its blowups.}

\section{Mirror manifolds, quantum corrections and discrete symmetries}

\subsection{Mirror manifolds from LG Superpotential}
In this section, we will follow \cite{HIV,AKV} to derive the
equations for the Calabi-Yau threefolds mirror to
local $\mathbb{P}^{2}$ blown up at three points and local
 $\mathbb{F}_{2}$ blown up at two points. The toric diagram for
these is shown in \figref{toric}.

\onefigure{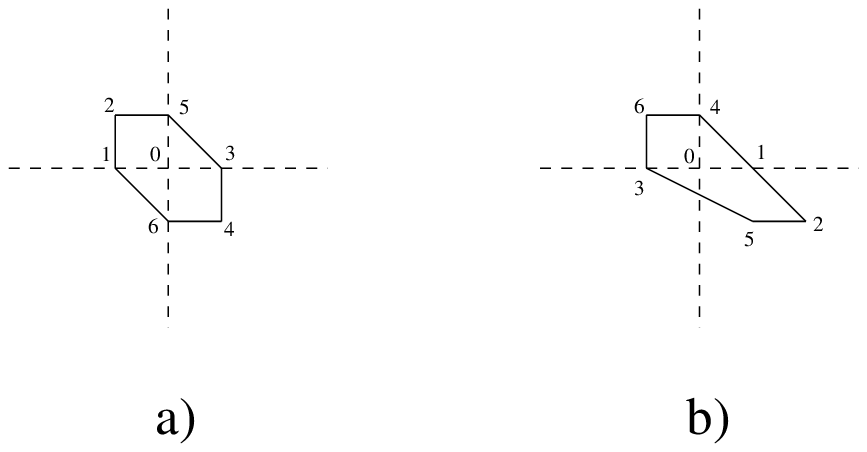}{a) $\mathbb{P}^{2}$ blown up at three points, b)
$\mathbb{F}_{2}$ blown up at two points. We will denote these surface
by ${\cal B}_{3}$ and ${\cal F}_{2}$ respectively. The numbers denote the order in which the vertices appear in the linear sigma model charge vectors.} 

{\bf Local ${\cal B}_{3}$:} Consider the non-compact CY which is the
total space of the anticanonical bundle over ${\cal B}_{3}$. The linear
sigma model charges for this Calabi-Yau space are given by \cite{YZL}
\bea l^{(1)}&=&(-2,1,0,1,0,0,0),\,\,\,l^{(2)}=(-2,0,1,0,1,0,0)\,, \label{morip} \\
\nonumber
l^{(3)}&=&(-1,1,-1,0,0,1,0),\,\,\,l^{(4)}=(-1,-1,1,0,0,0,1)\,. 
\eea The superpotential of the mirror Landau-Ginzburg theory is given
by \cite{HV}
\begin{eqnarray}
W &=& \sum_{i=0}^{6} e^{-Y_i}  \,.
\end{eqnarray}
The $Y_i$ are related to the chiral fields $\Phi_i$ of the linear sigma model via ${\mathrm Re}(Y_i)=|\Phi_i|^2$. The D-term constraints on the $\Phi_i$ at low energies, as encoded in the charges (\ref{morip}), translate into relations among the mirror variables $Y_i$. For ${\cal B}_{3}$, expressing the superpotential in terms of the independent variables $Y_0$, $Y_1$ and $Y_2$ yields
\bea
W(Y_{0},Y_{1},Y_{2})=x_{0}+x_{1}+x_{2}+e^{-t_{1}}\frac{x_{0}^{2}}{x_{1}}+e^{-t_{2}}\frac{x_{0}^{2}}{x_{2}}+e^{-t_{3}}\frac{x_{0}x_{2}}{x_{1}}+e^{-t_{4}}\frac{x_{0}x_{1}}{x_{2}}\,,
\eea
where $x_{i}=e^{-Y_{i}}$ and the $t_{i}$ are the K\"ahler parameters
of the original Calabi-Yau. The periods of the Landau-Ginzburg theory are
given by \cite{CV,HV,HIV} \bea \Pi=\int
e^{-W}\prod_{i}\frac{dx_{i}}{x_{i}}\,,  \eea where the measure of
integration reflects the fact that the $Y_{i}$ are the fundamental fields.
Because the superpotential is homogeneous of degree one, we can rewrite
the above integral in the following way, \bea \Pi&=&\int
e^{-x_{0}\{W(0,Y_{1}-Y_{0},Y_{2}-Y_{0})-zw\}}dx_{0}\,dz\,dw\,\frac{dx_{1}dx_{2}}{x_{1}x_{2}}
\\ \nonumber &=&\int \delta(W(0,Y_{1}-Y_{0},Y_{2}-Y_{0})-zw)
dz\,dw\frac{d\tilde{x}_{1}d\tilde{x}_{2}}{\tilde{x}_{1}\tilde{x}_{2}}\,,\,\,\tilde{x}_{i}=e^{-\tilde{Y}_{i}}=e^{-Y_{i}+Y_{0}}\,.
\eea Thus, the periods of the mirror LG theory are equivalent to the
integral of the holomorphic 3-form over a 3-cycle in a Calabi-Yau 3-fold
\cite{HIV}, \bea \Pi=\int \Omega \,,\,\,\,\,\,\,\Omega=dz dw
\frac{d\tilde{x}_{1}d\tilde{x}_{2}}{\tilde{x}_{1}\tilde{x}_{2}}/df\,,
\label{calabiyau}\eea
where $f=W(0,\tilde{Y}_{1},\tilde{Y}_{2})-zw$ and $f=0$ defines the Calabi-Yau.
The Calabi-Yau mirror to local ${\cal
B}_{3}$ is hence given by \bea
P_{{\cal B}_{3}}(u,v)=1+e^{u}+e^{v}+e^{-t_{1}-u}+e^{-t_{2}-v}+e^{-t_{3}-u+v}+e^{-t_{4}+u-v}=zw\,,
\eea where we have substituted $u,v$ for
$-\tilde{Y}_{1},-\tilde{Y}_{2}$ respectively.

{\bf Local ${\cal F}_{2}$:} Now consider the Calabi-Yau which is the
total space of the anticanonical bundle over $\mathbb{F}_{2}$ blown up at
two points. In this case, the linear sigma model charges are \bea
l^{(1)}&=&(-2,1,0,1,0,0,0)\,,\,\,l^{(2)}=(0,-2,1,0,1,0,0)\,,\\
\nonumber
l^{(3)}&=&(-1,1,-1,0,0,1,0)\,,\,\,l^{(4)}=(-1,-1,1,0,0,0,1)\,. \label{morif} \eea
From Eq(\ref{calabiyau}), it follows that the mirror Calabi-Yau in this
case is given by \bea P_{{\cal
F}_{2}}(u,v)=1+e^{u}+e^{v}+e^{-t_{1}-u}+e^{-t_{2}+2u-v}+e^{-t_{3}-u+v}+
e^{-t_{4}+u-v}=zw\,,
\eea and the holomorphic 3-form by \bea \Omega=dudv
\frac{dz}{z}\,.  \eea

\subsection{Open string topological amplitudes} 

Open
string topological amplitudes calculate certain superpotential terms
in the effective ${\cal N}=1$ theory on the world-volume of a D6-brane
wrapped on a Lagrangian 3-cycle of a Calabi-Yau threefold. These terms
in the effective theory are of the form \cite{BCOV,AKV} \bea h\int
d^{2}\theta (\mbox{Tr}W^{2})^{h-1}({\cal W}^{2})^{g}\,, \eea where $g$
is the genus, $h$ is the number of boundaries, $W$ is the gaugino
superfield and ${\cal W}$ is the ${\cal N}=2$ graviphoton multiplet.

The open string amplitudes get contributions from the BPS domain walls
in four dimensions, which correspond to D4-branes ending on the
D6-branes. These D4-branes are classified by the relative homology
classes in the CY3-fold with boundary on the Lagrangian cycle on which the
D6-brane is wrapped. It was shown in \cite{AV,AKV} that the Lagrangian
cycle in the A-model geometry maps to a holomorphic 2-cycle $\Sigma$
in the B-model geometry of the mirror Calabi-Yau. Hence, under mirror
symmetry, the D6-brane on a non-compact 3-cycle becomes a D5-brane
wrapped on a non-compact 2-cycle. This mirror D-brane is parametrized
by $z$ and its position is given by \bea w&=&0\,,\\ \nonumber P(u,v)&=&0\,.
\eea Thus, the holomorphic 2-cycle $\Sigma$, given by $P(u,v)=0$,
parameterizes the position of the mirror D-brane. The superpotential in
this case is given by \cite{AV,AKV} \bea W_{{\cal
N}=1}=\int_{u_{0}}^{u} v(u)du\,, \eea where $v(u)$ is obtained by
solving $P(u,v)=0$ and $u_{0}$ is a fixed point on $\Sigma$.

In order to obtain invariants associated with disks ending on the
Lagrangian cycle of the D6-brane, the above expression for the
superpotential has to be compared with the general form of the genus
zero amplitude predicted in \cite{OV}, \bea W_{{\cal N}=1}=\sum_{\Sigma\in
H_{2}(X,\bbbz)} \sum_{m\in \bbbz}\sum_{n=1}^{\infty}
\frac{N_{\Sigma,m}}{n^{2}}q^{n\Sigma}e^{nm\widehat{u}}\,.  \eea Here,
$q^{\Sigma}=e^{\int_{\Sigma} \omega}$ and $N_{\Sigma,m}$ is the number
of D4-branes which wrap the curve $\Sigma$ and end on the D6-brane by
winding $m$ times around the non-trivial 1-cycle of the Lagrangian
3-cycle on which the D6-brane is wrapped. In the above expression, $\omega$
is the quantum corrected K\"ahler form for the Calabi-Yau and
$\widehat{u}$ is the quantum corrected area of the disk.

\subsection{PF equations, mirror map and relations between 
quantum corrections} \label{mayr}

In this section, we show, based on a recent paper \cite{mayr} by Mayr,
how relations between the ratios $\log\frac{z_i}{q_i}$, where
$\mbox{log}(z_i)$ are the classical and $\mbox{log}(q_i)$ the quantum
corrected K\"ahler parameters, can be determined without solving the
Picard-Fuchs equations.

Consider a Calabi-Yau manifold with linear sigma model charges
$l^{(\alpha)}_{i}$. The Picard-Fuchs differential operators are then
given by \bea {\cal D}_{\alpha}= \prod_{l^{(\alpha)}_{i}>0}
(\frac{\partial}{\partial a_{i}})^{l^{(\alpha)}_{i}}
-\prod_{l^{(\alpha)}_{i}<0} (\frac{\partial}{\partial
a_{i}})^{-l^{(\alpha)}_{i}} \,, \eea where $a_{i}$ are complex
structure parameters, but not all of them are independent. Only the
following combinations lead to independent complex structure
parameters of the mirror Calabi-Yau, \bea
z_{\alpha}=\prod_{i}a_{i}^{l^{(\alpha)}_{i}}\,.\eea Here the number of
$z_{\alpha}$ is equal to the number of K\"ahler parameters of the
original Calabi-Yau. In terms of the $z_{\alpha}$ the above
differential operators are given by \cite{mayr} \bea {\cal
D}_{\alpha}=\prod_{l^{(\alpha)}_{i}>0}\prod_{j=0}^{l^{(\alpha)}_{i}-1}(
\sum_{\beta}l^{(\beta)}_{i}\theta_{\beta}-j)-z_{\alpha}\prod_{l^{(\alpha)}_{i}<0}\prod_{j=0}^{-l^{(\alpha)}_i-1}
(\sum_{\beta}l^{(\beta)}_{i}\theta_{\beta}-j)\,, \label{PF} \eea where
$\theta_{\beta}=z_{\beta}\frac{\partial}{\partial z_{\beta}}$.
Using the quantum corrected K\"ahler parameter
$-\mbox{log}(q)$ (the solution of the PF equations) we define \bea
R_{\beta}=\mbox{log}(z_{\beta}/q_{\beta})\,. \eea The $R_{\beta}$
satisfy the following differential equation \cite{mayr}, \bea {\cal
D}_{\alpha} R_{\beta}=z_{\alpha}A^{\alpha}_{\beta}\,, \eea where the
$A^{\alpha}_{\beta}$ are defined by the linear part of the
differential operator ${\cal D}_{\alpha}$ \cite{mayr}, ${\cal D}_{\alpha}^{\mathrm lin}=z_{\alpha} \sum A^{\alpha}_{\beta} \theta_\beta$. We have here used the fact that if there is more than one
positive number in the vector $l^{(\alpha)}$, as is the case in all the examples we will be considering, then the first term in Eq(\ref{PF}) does not contribute to ${\cal D}_{\alpha}^{\mathrm lin}$. If in addition, there is more than one negative number, then $ A^{\alpha}_{\beta}$ vanishes. If the vector $l^{(\alpha)}$ has only one
negative number, say $l_{s}^{(\alpha)}$, then \bea
A^{\alpha}_{\beta}=(\prod_{j=1}^{-l^{(\alpha)}_{s}-1}j)\,l^{(\beta)}_{s}\,. \label{lind}
\eea

Thus, we see that if $\sum_{\beta}a_{\beta}A^{\alpha}_{\beta}=0$ for
all $\alpha$, \bea {\cal
D}_{\alpha}(\sum_{\beta}a_{\beta}R_{\beta})=0 \,\,\,,\forall \,\alpha\,.\eea Since $R_{\beta}$
is a power series in $z_{\alpha}$, we get the following
relation\,, \bea \sum_{\beta}a_{\beta}
R_{\beta}=0\,.  \label{relation}\eea We use this relation to verify
that the quantum corrections to the boundary variables obtained using
discrete symmetries of the superpotential do not lead to inconsistent
results.

Let's apply this method to our two examples.
The charges for local ${\cal B}_{3}$ are given in Eq(\ref{morip}). From Eq(\ref{lind}), it then follows that \bea
A^{1}_{\beta}=A^{2}_{\beta}=l^{(\beta)}_{1}=
(-2,-2,-1,-1)\,,\,\,A^{3}_{\beta}=A^{4}_{\beta}=0\,.  \eea
Therefore, the solution to the equation $\sum_{\beta}a_{\beta}A^{\alpha}_{\beta}=0$ is given by $a_{\beta}=(a_{1},a_{2},a_{3},-2a_{1}-2a_{2}-a_{3})$ yielding the following relations,
\bea z_{1}=q_{1}(\frac{z_{4}}{q_{4}})^{2}\,,\,\,
z_{2}=q_{2}(\frac{z_{4}}{q_{4}})^{2}\,,\,\,
z_{3}=q_{3}(\frac{z_{4}}{q_{4}})\,.  \eea

For local ${\cal F}_{2}$, we get, using the charges (\ref{morif}),
\begin{eqnarray}
A^{1}_{\beta}&=&(-2,0,-1,-1) \,, \\ \nonumber
A^{2}_{\beta}&=&(1,-2,1,-1) \,, \\ \nonumber
A^{3}_{\beta}&=& A^{4}_{\beta} \,\, = \,\, 0  \,.
\end{eqnarray}

We now solve   $\sum_{\beta}a_{\beta}A^{\alpha}_{\beta}=0$ for all $\alpha$ by $a_{\beta}=(a_{1},-\frac{a_{1}+2a_4}{2},-2a_{1}-a_{4},a_{4})$, yielding the two relations
\begin{eqnarray}
z_1=q_1 \sqrt{\frac{z_2}{q_2}}(\frac{z_3}{q_3})^2 \,,\,\, z_4=q_4 \frac{z_2}{q_2} \frac{z_3}{q_3} \,. \label{mayrf}
\end{eqnarray}

We will obtain these relations again using the discrete symmetries of the superpotential, and ultimately be able to verify them by considering the explicit logarithmic solutions to the PF equations.

\subsection{Discrete symmetries of the LG superpotential and 
quantum corrected variables} Consider the LG theory which is the
 mirror of the Calabi-Yau twofold ${\cal O}(-2) \mapsto
 \mathbb{P}^{1}$. The superpotential is given by \bea
 W(\Sigma,Y_{0},Y_{1},Y_{2})=\Sigma(Y_{1}+Y_{2}-2Y_{0}-t)+e^{-Y_{0}}
 +e^{-Y_{1}}+e^{-Y_{2}}\,.  \eea We see that there is a $\mathbb{Z}_{2}$
 symmetry which interchanges $Y_{1}$ and $Y_{2}$. After integrating
 out $\Sigma$, we get \bea
 W=e^{-Y_{0}}+e^{-Y_{1}}+e^{Y_{1}-2Y_{0}-t}\,.  \eea Now, the
 $\mathbb{Z}_{2}$ symmetry corresponds to \bea e^{-Y_{1}} &\mapsto&
 e^{Y_{1}-2Y_{0}-t}\,,\\ \nonumber x_{1}&\mapsto&
 \frac{x_{0}^{2}e^{-t}}{x_{1}}\,,\,\,\,\,x_{i}=e^{-Y_{i}}.  \eea
 Although this model does not correspond to a Calabi-Yau threefold, we
 can obtain a similar superpotential from a degenerate limit of the
 superpotential mirror to the local $\mathbb{P}^{1}\times
 \mathbb{P}^{1}$. Here, we are just using this superpotential as an example to 
illustrate the method we will use later.

Let us define quantum corrected variables \bea
\widehat{x}_{i}=S_{i}(q)x_{i}\,,\,\,\,q=e^{-T}\,,  \eea where $T$ is
the quantum corrected area of the $\mathbb{P}^{1}$.
The superpotential in terms of these new variables is given by
\bea
W=\frac{\widehat{x}_{0}}{S_{0}(q)}+\frac{\widehat{x}_{1}}{S_{1}(q)}+
e^{-t}\frac{S_{1}}{S^{2}_{0}(q)}\frac{\widehat{x}^{2}_{0}}{\widehat{x}_{1}}\,.
\eea
If we now assume that the $\mathbb{Z}_{2}$ symmetry of the superpotential still exists in terms of the new variables, i.e.
\bea
\widehat{x}_{1}\mapsto e^{-T}\frac{\widehat{x}^{2}_{0}}{\widehat{x}_{1}}\,,
\eea
we get the following relation between $S_{0}(q)$ and $S_{1}(q)$,
\bea
\frac{S_{1}(q)}{S_{0}(q)}=\sqrt{\frac{q}{z}}\,,\,\,z=e^{-t}\,.
\eea

We will use this basic $x\mapsto 1/x$ symmetry for the Calabi-Yau
threefold cases to compute quantum corrections.

\subsubsection{Examples}In this subsection, we derive the quantum corrections 
to the boundary variables for certain cases discussed in \cite{AKV} using the
$\mathbb{Z}_{2}$ symmetry of the superpotential.

\underline{{\bf ${\cal O}(-3)$ over $\mathbb{P}^{2}$:}} The superpotential of the
mirror theory is given by \cite{HV}\bea
W=x_{0}+x_{1}+x_{2}+e^{-t}\frac{x_{0}^{3}}{x_{1}x_{2}} \,.\eea Let 
the quantum corrected variables be
\bea
\widehat{x}_{i}=\frac{x_{i}}{S_{i}(q)}\,.
\eea
Since $x_{1}$ and $x_{2}$ can be interchanged without affecting the
superpotential, $S_{1}(q)=S_{2}(q)$. Also, we see that the following transformation leaves the superpotential invariant,
\bea
x_{1}\mapsto e^{-t}\frac{x_{0}^{3}}{x_{1}x_{2}}\,.
\eea
Assuming this symmetry holds in terms of the quantum corrected variables, i.e.
\bea
\widehat{x}_{1}\mapsto e^{-T}\frac{\widehat{x}^{3}_{0}}{\widehat{x}_{1}\widehat{x}_{2}}\,,
\eea
implies that  \bea
\frac{S_{1}(q)}{S_{0}(q)}=(\frac{q}{z})^{\frac{1}{3}}\,.  
\eea
In the phase in which $x_{0}=1$ and therefore $S_{0}(q)=1$, the above equation
gives the same result as in \cite{AKV}. 

\underline{{\bf local $\mathbb{P}^{1}\times \mathbb{P}^{1}$:}} In this case, the
superpotential of the mirror LG theory is given by \cite{HV,HIV} \bea
W=x_{0}+x_{1}+x_{2}+e^{-t_{1}}\frac{x_{0}^{2}}{x_{1}}+e^{-t_{2}}\frac{x_{0}^{2}}{x_{2}}\,.
\eea

The transformations \bea x_{1}\mapsto
e^{-t_{1}}\frac{x^{2}_{0}}{x_{1}}\,,\,\,\, x_{2}\mapsto
e^{-t_{2}}\frac{x^{2}_{0}}{x_{2}}  \eea implemented in terms of the
quantum corrected variables $\widehat{x}_{i}=S_{i}(q_{1},q_{2})x_{i}$ read
\begin{eqnarray}
 \widehat{x}_{1}\mapsto
e^{-T_{1}}\frac{\widehat{x}^{2}_{0}}{\widehat{x}_{1}}\,,\,\,\, \widehat{x}_{2}\mapsto
e^{-T_{2}}\frac{\widehat{x}^{2}_{0}}{\widehat{x}_{2}}\,.
\end{eqnarray}
Invariance under this transformation implies \bea S_{1}(q_{1},q_{2})&=&\sqrt{\frac{q_{1}}{z_{1}}}\,,\,\,\, S_{2}(q_{1},q_{2})=\sqrt{\frac{q_{2}}{z_{2}}} \,.\eea

\underline{{\bf local $\mathbb{P}^{2}$ blown up at one point:}}

The superpotential of the mirror LG theory is given by \cite{HV,AKV} \bea
W=x_{0}+x_{1}+x_{2}+e^{-t_{b}-t_{f}}\frac{x_{0}^{3}}{x_{1}x_{2}}+e^{-t_{f}}\frac{x_{0}^{2}}{x_{2}}\,.
\eea
The superpotential is invariant under the transformation
\bea
x_{2}\mapsto \frac{1}{x_{2}}(z_{b}z_{f}\frac{x_{0}^{3}}{x_{1}}+z_{f}x_{0}^{2})\,.
\eea
In terms of quantum corrected variables $\widehat{x}_{i}=S_{i}(q_{b},q_{f})x_{i}$ we require that the superpotential be invariant under the transformation,
\bea
\widehat{x}_{2} \mapsto \frac{1}{\widehat{x}_{2}}(q_{b}q_{f}\frac{\widehat{x}_{0}^{3}}{\widehat{x}_{1}}+q_{f}\widehat{x}_{0}^{2})\,.
\eea
This requirement uniquely fixes $S_{i}$ to be
\bea
\frac{S_{1}(q_{b},q_{f})}{S_{0}(q_{b},q_{f})}=\frac{q_{b}}{z_{b}}\,,\,\,\,\frac{S_{2}(q_{b},q_{f})}{S_{0}(q_{b},q_{f})}=\sqrt{\frac{q_{f}}{z_{f}}}\,.
\eea
In the phase $x_0=1$, $S_0=1$, and the above relations determine $S_1$ and $S_2$.
Using the relation $\frac{q_{f}}{z_{f}}=(\frac{q_{b}}{z_{b}})^{2}$, derived eg. from
Eq(\ref{relation}), we see that $S_{1}=S_{2}$.

\section{Integrality of Ooguri-Vafa open string invariants} 
\subsection{Local ${\cal F}_{2}$}

The linear sigma model charges for this case are \cite{YZL}
\begin{eqnarray}
l^{(1)}&=& (-2,1,0,1,0,0,0) \,,\,\, l^{(2)}= (0,-2,1,0,1,0,0) \,, \\ \nonumber
l^{(3)}&=& (-1,1,-1,0,0,1,0) \,,\,\, l^{(4)}= (-1,-1,1,0,0,0,1) \,.
\end{eqnarray}

The mirror to this theory is a Landau-Ginzburg model with superpotential 
\cite{HV}
\begin{eqnarray}
W = x_0 + x_1 + x_2 + z_1 \frac{x_0^2}{x_1} + z_2 \frac{x_1^2}{x_2} + z_3 \frac{x_0 x_2}{x_1} + z_4 \frac{x_0 x_1}{x_2} \,,\,\,\,z_{i}=e^{-t_{i}}\,.
\end{eqnarray}

The above superpotential is invariant under the transformation \bea
x_{2}\mapsto
\frac{1}{x_{2}}(\frac{z_{2}x_{1}^{2}+z_{4}x_{0}x_{1}}{1+z_{3}\frac{x_{0}}{x_{1}}})\,.
\eea In terms of the quantum corrected variables
$\widehat{x}_{i}=S_{i}x_{i}$, we require that the superpotential be
invariant under the transformation, \bea \widehat{x}_{2}\mapsto
\frac{1}{\widehat{x}_{2}}(\frac{q_{2}\widehat{x}^{2}_{1}+q_{4}\widehat{x}_{0}\widehat{x}_{1}}{1+q_{3}\frac{\widehat{x}_{0}}{\widehat{x}_{1}}})\,.
\eea This implies that \bea
\frac{S_{1}}{S_{0}}=\frac{q_{3}}{z_{3}}\,,\,\,\frac{S_{2}}{S_{0}}=\sqrt{\frac{q_{4}q_{3}}{z_{4}z_{3}}}\,.
\eea It also yields the following relation, \bea
\frac{q_{4}}{z_{4}}=\frac{q_{2}}{z_{2}}\frac{q_{3}}{z_{3}}\,,  \eea
consistent with the result (\ref{mayrf}) derived above.

We consider the phase in which the $x_{0}=1$ and set
$\widehat{x}_{1}=-e^{\widehat{u}},
\widehat{x}_{2}=-e^{\widehat{v}}$ \cite{AKV}. The equation parameterizing the
position of the brane in the mirror geometry that we must solve is
\begin{equation} 
1-\frac{e^{\widehat{u}}}{S_{1}}-\frac{e^{\widehat{v}}}{S_{2}}-S_{1}e^{-t_{1}-\widehat{u}}-\frac{S_{2}}{S^{2}_{1}}e^{-t_{2}+2\widehat{u}-\widehat{v}}+\frac{S_{1}}{S_{2}}e^{-t_{3}+\widehat{v}-\widehat{u}}+
\frac{S_{2}}{S_{1}}e^{-t_{4}+\widehat{u}-\widehat{v}} = 0 \,.
\label{mirror}
\end{equation}

The logarithmic solutions to the PF equations have the general form\footnote{In applying these formulae, the identity $\frac{d}{dx}\frac{1}{\Gamma(x)} \Big{|}_{x=1-n}=(-1)^{n+1} \Gamma(n)$ for n a positive integer proves useful.}
\bea
\Pi_i &=& \partial_{\rho_i} \Pi_0 \big{|}_{\vec{\rho}=0}\,,
\eea
where
\bea \label{solpf}
\Pi_0(\vec{z})&=&\sum_{\vec{n}} c(\vec{n},\vec{\rho}) \vec{z}^{\,\vec{n}+\vec{\rho}}\big{|}_{\vec{\rho}=0}\,,\\ \nonumber
c(\vec{n},\vec{\rho})&=&\frac{1}{\prod_i \Gamma(\sum_{\alpha} l_i^{\alpha}(n_\alpha + \rho_\alpha)+1)}\,. 
\eea

For the case we are considering, these solutions 
can be expressed in terms of
\bea
A&=&\sum_{m,n,p,r}\frac{(-1)^{p+r}\Gamma(2m + p + r)}{\Gamma(m -2n +p-r+1)\Gamma(n-p+r+1)m!n!p!r!}z_1^m z_2^n z_3^p z_4^r \nonumber \\
B&=&\sum_n \frac{\Gamma(2n)}{n!^2}z_2^n 
\eea
as
\begin{eqnarray}
-T_{1}=\log(q_1)&=& \log(z_1)+2A-B \,, \\ \nonumber
-T_{2}=\log(q_2)&=& \log(z_2)+2B \,, \\ \nonumber
-T_{3}=\log(q_3)&=& \log(z_3)+A-B \,, \\ \nonumber
-T_{4}=\log(q_4)&=& \log(z_4)+A+B \,.
\end{eqnarray}
To express the solution to Eq(\ref{mirror}) purely in terms of B-model variables, we need to invert the mirror map,
\begin{eqnarray}
z_1 &=& q_1 (1-2q_1+q_2+3q_1^2-4 q_1 q_2 +4 q_1 q_4 - 2 q_3 q_4 + \ldots) \,, \nonumber  \\
z_2 &=& \frac{q_2}{(1+q_2)^2} \,;
\end{eqnarray}
$z_3$ and $z_4$ can be obtained from the above and the relations among the parameters.

Of the two solutions for $e^{\widehat{v}}$, we choose the one that
reduces to $1-e^{\widehat{u}}$ in the large radius limit
$z_{\alpha} \rightarrow 0$, since this is the relation we know to hold classically,
\bea \nonumber
\widehat{v}=\mbox{log}[\frac{S_{2}e^{\widehat{u}}-\frac{S_{2}}{S_{1}}e^{2\widehat{u}}-S_{1}S_{2}z_{1}+\sqrt{(S_{2}x-\frac{S_{2}}{S_{1}}e^{2\widehat{u}}-S_{1}S_{2}z_{1})^{2}-4(\frac{e^{\widehat{u}}}{S_{1}}-z_{3})(S_{2}^{2}z_{2}e^{3\widehat{u}}z_{2}+S_{2}^{2}z_{4}e^{2\widehat{u}})}}{2(e^{\widehat{u}}-S_{1}z_{3})}]\,.
\eea

Having expressed $\hv$ completely in terms of flat coordinates of the B-model
variables, we are now ready to calculate the instanton numbers
$N_{\vec{k},m}$ using the relation \cite{AV,AKV}
\begin{eqnarray}
\widehat{v}=\sum_{\vec{k}\in H_{2}(X,\mathbb{Z})}\sum_{m\in \mathbb{Z}}\sum_{n=1}^{\infty}
\frac{m}{n}N_{\vec{k},m}q_{1}^{nk_{1}}q_{2}^{nk_{2}}q_{3}^{nk_{3}}q_{4}^{nk_{4}}e^{nm\widehat{u}}\,.
\end{eqnarray}
Below, we tabulate the functions $I_{\vec{k}}(\hx)$, which we define as

\begin{eqnarray}
I_{\vec{k}}(\hx) &=& \sum_{m\neq 0} N_{\vec{k},m}\,\hx^m \,.
\end{eqnarray}
The $N_{\vec{k},m}$ can easily be determined from these via Taylor
expansion, and pass the integrality test. We give the $I_{\vec{k}}(\hx)$ for 
$\sum_{i=1}^{4}k_{i}=1,2$ below and for $\sum_{i=1}^{4}k_{i}=3,4,5$ in the appendix.

\begin{eqnarray}
\begin{array}{|c|c||c|c|}
\hline 
\rule{0mm}{6mm}\vec{k}& I_{\vec{k}}(x)&\vec{k} & I_{\vec{k}}(x) \\ \hline \hline
\rule{0mm}{6mm}(1000)& \frac{1}{x}&(0002) &-\frac{x^2}{(1+x)(1-x)^3}\\ \hline 
\rule{0mm}{6mm}(0100) & -\frac{x}{1-x} &(1100) & -\frac{2x}{1-x}    \\ \hline
\rule{0mm}{6mm}(0010) & -\frac{1}{x}&   (1010) &    0     \\ \hline  
\rule{0mm}{6mm}(0001) & \frac{x}{1-x}& (1001) &    \frac{2x}{1-x}  \\ \hline 
\rule{0mm}{6mm}(2000) & 0 &  (0110) &     \frac{x}{1-x}     \\ \hline
\rule{0mm}{6mm}(0200) &    -\frac{x^3}{(1-x)^3(1+x)}&  (0101) &      \frac{x^2}{(1-x)^3}   \\ \hline
\rule{0mm}{6mm}(0020) &       0 & (0011) &    -\frac{x}{1-x}   \\ \hline \hline
\end{array}
\nonumber
\end{eqnarray}

\subsection{Local ${\cal B}_{3}$}
Here we consider the non-compact Calabi-Yau which is the total space of
the anticanonical bundle on toric del Pezzo ${\cal B}_{3}$. The linear 
sigma model with moduli space this Calabi-Yau has the charge vectors \bea
l^{(1)}&=&(-2,1,0,1,0,0,0)\,,\,\,l^{(2)}=(-2,0,1,0,1,0,0)\,,\\ \nonumber
l^{(3)}&=&(-1,1,-1,0,0,1,0)\,,\,\,l^{(4)}=(-1,-1,1,0,0,0,1)\,.  \eea
The superpotential of the mirror theory is given by \bea
W=x_{0}+x_{1}+x_{2}+e^{-t_{1}}\frac{x^{2}_{0}}{x_{1}}
+e^{-t_{2}}\frac{x^{2}_{0}}{x_{2}}+e^{-t_{3}}
\frac{x_{0}x_{2}}{x_{1}}+e^{-t_{4}}\frac{x_{0}x_{1}}{x_{2}}\,.  \eea
The above superpotential is invariant under the transformation \bea
x_{1}\mapsto
\frac{1}{x_{1}}(\frac{z_{1}x^{2}_{0}+z_{3}x_{0}x_{2}}{1+z_{3}\frac{x_{0}}{x_{2}}})\,.
\eea In terms of the quantum corrected variables
$\widehat{x}_{i}=S_{i}(q)x_{i}$ we require \bea
\widehat{x}_{i}\mapsto
\frac{1}{\widehat{x}_{2}}(\frac{q_{1}\widehat{x}^{2}_{0}+q_{3}\widehat{x}_{0}\widehat{x}_{2}}{1+q_{4}\frac{\widehat{x}_{0}}{\widehat{x}_{2}}})\,.
\eea
This implies that 
\bea
\frac{S_{1}}{S_{0}}=\sqrt{\frac{q_{1}}{z_{1}}}\,,\,\,\,\frac{S_{2}}{S_{0}}=\frac{q_{4}}{z_{4}}\,,
\eea and we also get a relation \bea \frac{q_{1}}{z_{1}}
=\frac{q_{4}}{z_{4}}\frac{q_{3}}{z_{3}}\,. \eea This relation agrees with those determined in section \ref{mayr}.
Using these relations, we only need to determine $z_{4}$ in terms of the $q_{\alpha}$.

The general formula (\ref{solpf}) allows us to express the logarithmic solutions to the PF equations as

\begin{eqnarray}
\log(q_1)&=& \log(z_1)+2A \,, \\ \nonumber
\log(q_2)&=& \log(z_2)+2A \,, \\ \nonumber
\log(q_3)&=& \log(z_3)+A \,, \\ \nonumber
\log(q_4)&=& \log(z_4)+A \,,
\end{eqnarray}

where $A$ is given by
\begin{eqnarray}
A&=& \sum_{m,n,p,r} \frac{(-1)^{p+r} \Gamma(2m+2n+p+r)}{\Gamma(m+p-r+1)\Gamma(n-p+r-1)m!n!p!r!} \,.
\end{eqnarray}

 By inverting
the solution $\mbox{log}(z_{4})+A$ of the PF equation, we get
\bea
z_{4}=q_{4}(1-q_{1}+q_{1}^{2}-q_{1}^{3}+q_{1}^{4}-q_{2}-q_{1}q_{2}
-2q_{1}^{2}q_{2}-2q_{1}^{3}q_{2}-3q_{1}^{4}q_{2}+q_{2}^{2}+\cdots)\,.
\eea
In the phase in which $x_{0}=1$, the Riemann surface parameterizing the
position of the mirror brane is given by \bea
S_{1}(q)-e^{\widehat{u}}-e^{\widehat{v}}-q_{1}e^{-\widehat{u}}-q_{2}e^{-\widehat{v}}+q_{3}e^{\widehat{v}-\widehat{u}}
+q_{4}e^{\widehat{u}-\widehat{v}}=0\,,\,\,\widehat{x}_{1}=-e^{\widehat{u}},
\widehat{x}_{2}=-e^{\widehat{v}}\,. \eea
Solving for $\widehat{v}$ yields 
\bea
\widehat{v}=\mbox{log}\frac{S_{1}(q)-e^{\widehat{u}}-q_{1}e^{-\widehat{u}}+\sqrt{(S_{1}(q)-e^{\widehat{u}}-q_{1}e^{-\widehat{u}})^{2}-4(1-q_{3}e^{-\widehat{u}})(q_{2}-q_{4}e^{\widehat{u}})}}{2(1-q_{3}e^{-\widehat{u}})}\,.
\eea
To obtain the disk invariants, we compare this solution with 
\bea
\widehat{v}=\sum_{\vec{k}\in H_{2}(X,\mathbb{Z})}\sum_{m\in \mathbb{Z}}\sum_{n=1}^{\infty}
\frac{m}{n}N_{\vec{k},m}q_{1}^{nk_{1}}q_{2}^{nk_{2}}q_{3}^{nk_{3}}q_{4}^{nk_{4}}e^{nm\widehat{u}}\,,
\eea
and define as before, \bea I_{\vec{k}}(\hx)=\sum_{m \neq
0}N_{\vec{k},m}\hx^{m}\,.  \eea The functions $I_{\vec{k}}(x)$ for
$\sum_{i=1}^{4}k_{i}=1,2$ are given in the table below.  The results for
$\sum_{i=1}^{4}k_{i}=3,4,5$ are given in the appendix.
\begin{eqnarray}
\begin{array}{|c|c||c|c|}
\hline 
\rule{0mm}{6mm}\vec{k}  & I_{\vec{k}}(x)& \vec{k} &I_{\vec{k}}(x)\\\hline\hline
\rule{0mm}{6mm}(1000)& \frac{1}{x}& (0002) &-\frac{x^{2}}{(1+x)(1-x)^3}  \\ \hline 
\rule{0mm}{6mm}(0100)& -\frac{x}{1-x}&(1100)&\frac{1}{x}-\frac{2x}{1-x} \\ \hline
\rule{0mm}{6mm}(0010)& -\frac{1}{x}& (1010) &0   \\ \hline  
\rule{0mm}{6mm}(0001)& \frac{x}{1-x}&(1001) &\frac{2x}{1-x}    \\ \hline 
\rule{0mm}{6mm} (2000) & 0 & (0110) &-\frac{1}{x}+\frac{x}{1-x}  \\ \hline
\rule{0mm}{6mm}(0200)&-\frac{x}{(1+x)(1-x)^3}  &  (0101) & \frac{x}{(1-x)^3}\\ \hline 
\rule{0mm}{6mm} (0020) & 0 & (0011) &-\frac{x}{1-x} \\ \hline   
\end{array}
\nonumber
\end{eqnarray}

\section*{Acknowledgements}
A.I. would like to thank Julie D. Blum for valuable discussions.
This research was supported in part by NSF grant PHY-0071512.

\appendix
\section{Local ${\cal F}_{2}$}

\begin{eqnarray}
\begin{array}{|c|c||c|c||c|c|}
\hline 
\rule{0mm}{6mm} \vec{k}& I_{\vec{k}}(x) &\vec{k} &I_{\vec{k}}(x)& \vec{k} &I_{\vec{k}}(x)\\ \hline \hline
\rule{0mm}{6mm} (3000) &   0  &(2100) &    -\frac{3x}{1-x}& (2010) &     0   \\ \hline \hline
\rule{0mm}{6mm} (2001) &   \frac{3x}{1-x}-\frac{1}{x} & (1200) &  -\frac{2x^2}{(1-x)^3} & (1110) & \frac{2x}{1-x}  \\ \hline \hline
\rule{0mm}{6mm} (1101) &   \frac{2x(1+x)}{(1-x)^3} & (1020) &      0  & (1011) &   \frac{1-x-2x^2}{x(1-x)}  \\ \hline \hline
\rule{0mm}{6mm} (1002) &   -\frac{2x}{(1-x)^3} & (0300) &    -\frac{x^4(1+x^2)}{(1-x)^5 (1+x+x^2)}  & (0210) &     \frac{x^2}{(1-x)^3}  \\ \hline \hline
\rule{0mm}{6mm} (0201) &      \frac{x^3(1+x)}{(1-x)^5} & (0120) &       0 &(0111) &     -\frac{x(1+x)}{(1-x)^3}  \\ \hline \hline
\rule{0mm}{6mm} (0102) &   -\frac{2x^3}{(1-x)^5} & (0030) &      0  & (0021) &      0  \\ \hline \hline
\rule{0mm}{6mm} (0012) &     \frac{x}{(1-x)^3}  & (0003) &      \frac{x^3(1+x)}{(1-x)^5(1+x+x^2)} &&  \\ \hline \hline
\end{array}
\nonumber
\end{eqnarray}

\begin{eqnarray}
\begin{array}{|c|c||c|c||c|c|}
\hline 
\rule{0mm}{6mm} \vec{k}  & I_{\vec{k}}(x) & \vec{k} & I_{\vec{k}}(x) &\vec{k} &I_{\vec{k}}(x) \\ \hline \hline
\rule{0mm}{6mm} (4000) &  0  & (3100) &  -\frac{4x}{1-x}+\frac{1}{x} & (3010) &  0    \\ \hline \hline
\rule{0mm}{6mm} (3001) &  -\frac{1+x-2x^2-4x^3}{(1-x)x^2} & (2200) &  -\frac{x(3+4x+3x^2)}{(1-x)^3 (1+x)}  & (2110) &  \frac{3x}{1-x}-\frac{1}{x}    \\ \hline \hline
\rule{0mm}{6mm} (2101) &  \frac{2x(9-8x+4x^2)}{(1-x)^3}  & (2020) &  0 & (2011) &  -\frac{3x}{1-x}+\frac{1}{x^2}+\frac{2}{x} \\ \hline \hline
\rule{0mm}{6mm} (2002) &  -\frac{x(15-2x-11x^2+8x^3)}{(1-x)^3 (1+x)}  & (1300) &  -\frac{2x^3 (1+x)}{(1-x)^5}  & (1210) &  \frac{2x(1+x)}{(1-x)^3}    \\ \hline \hline
\rule{0mm}{6mm} (1201) &  \frac{2x^2(1+4x+x^2)}{(1-x)^5}  & (1120) &  0  & (1111) &  -\frac{2x(7-6x+3x^2)}{(1-x)^3}    \\ \hline \hline
\rule{0mm}{6mm} (1102) &  -\frac{6x^2(1+x)}{(1-x)^5}   & (1030) &  0  & (1021) &  0    \\ \hline \hline
\end{array}
\nonumber
\end{eqnarray}

\begin{eqnarray}
\begin{array}{|c|c||c|c||c|c|}
\hline 
\rule{0mm}{6mm} \vec{k}      & I_{\vec{k}}(x)&\vec{k} & I_{\vec{k}}(x) &\vec{k}&I_{\vec{k}}(x) \\ \hline \hline
\rule{0mm}{6mm} (1012) &  \frac{2x(6-7x+3x^2)}{(1-x)^3} & (1003) &  \frac{4x^2}{(1-x)^5}  & (0400) &  -\frac{x^5 (1+2x +4x^2+2x^3+x^4)}{(1-x)^7 (1+x)^3}    \\ \hline \hline
\rule{0mm}{6mm} (0310) &  \frac{x^3 (1+x)}{(1-x)^5} & (0301) &  \frac{x^4 (1+3x+x^2)}{(1-x)^7}   & (0220) &  -\frac{x^2}{(1-x)^3 (1+x)}    \\ \hline \hline
\rule{0mm}{6mm} (0211) &  -\frac{x^2 (1+4x+x^2)}{(1-x)^5}  & (0202) &  -\frac{2x^4 (2+7x +12x^2+7x^3+2x^4)}{(1-x)^7 (1+x)^3} & (0130) &  0    \\ \hline \hline
\rule{0mm}{6mm} (0121) &  \frac{x}{(1-x)^3}& (0112) &  \frac{3x^2 (1+x)}{(1-x)^5}  & (0103) &  \frac{5x^4}{(1-x)^7}    \\ \hline \hline
\rule{0mm}{6mm} (0040) &  0  & (0031) &  0   & (0022) &  -\frac{x}{(1-x)^3 (1+x)}   \\ \hline \hline
\rule{0mm}{6mm} (0013) &  -\frac{2x^2}{(1-x)^5} & (0004) &  -\frac{2 x^4(1+3x+x^2)}{(1-x)^7 (1+x)^3}   & &  \\ \hline \hline
\end{array}
\nonumber
\end{eqnarray}

\begin{eqnarray}
\begin{array}{|c|c||c|c||c|c|}
\hline 
\rule{0mm}{6mm} \vec{k}      & I_{\vec{k}}(x) & \vec{k} & I_{\vec{k}}(x) & \vec{k} & I_{\vec{K}}(x)\\ \hline \hline
\rule{0mm}{6mm} (5000) &   0  & (4100) &   -\frac{5x}{1-x}+\frac{1}{x^2}+\frac{2}{x} & (4010) &   0    \\ \hline \hline
\rule{0mm}{6mm} (4001) &   -\frac{1+x+x^2-3x^3-5x^4}{(1-x)x^3}  & (3200) &   -\frac{10(2-2x+x^2)}{(1-x)^3} & (3110) &   -\frac{1+x-2x^2-4x^3}{(1-x)x^2}    \\ \hline \hline
\rule{0mm}{6mm} (3101) &   -2\frac{1-3x-37x^2+54x^3-25x^4}{(1-x)^3 x}  &(3020) &   0   & (3011) &   \frac{1+x+x^2-3x^3-4x^4}{x^3-x^4}    \\ \hline \hline
\rule{0mm}{6mm} (3002) &   5(-\frac{2}{(1-x)^3} - \frac{6}{1-x}+\frac{1}{x}+8)  & (2300) &   -\frac{x^2(3+8x+3x^2)}{(1-x)^5}   & (2210) &   \frac{2x(9-8x+4x^2)}{(1-x)^3}    \\ \hline \hline
\rule{0mm}{6mm} (2201) &   \frac{x(3+23x+8x^2+8x^3)}{(1-x)^5}  & (2120) &   0   & (2111) &   2\frac{1-3x-4x^2(9-13x+6x^2)}{(1-x)^3 x}   \\ \hline \hline
\rule{0mm}{6mm} (2102) &   -\frac{x(15+17x+5x^2+5x^3)}{(1-x)^5} & (2030) &   0  &&  \\ \hline \hline

\end{array}
\nonumber
\end{eqnarray}

\begin{eqnarray}
\begin{array}{|c|c||c|c||c|c|}
\hline 
\rule{0mm}{6mm} \vec{k}      & I_{\vec{k}}(x) &\vec{k} & I_{\vec{k}}(x) &\vec{k} &I_{\vec{k}}(x)\\ \hline \hline
\rule{0mm}{6mm} (2021) &   0  & (2012) &   -\frac{6}{x}+\frac{10x(6-9x+4x^2)}{(1-x)^3} & (2003) &   \frac{x(12-3x+5x^2)}{(1-x)^5}    \\ \hline \hline
\rule{0mm}{6mm} (1400) &   -\frac{2x^4(1+3x+x^2)}{(1-x)^7}   & (1310) &   \frac{2x^2 (1+4x+x^2)}{(1-x)^5}  & (1301) &   \frac{2x^3 (1+9x+9x^2+x^3)}{(1-x)^7}   \\ \hline \hline
\rule{0mm}{6mm} (1220) &   -\frac{2x}{(1-x)^3}    & (1211) &   -\frac{2x (1+10x+4x^2+3x^3)}{(1-x)^5}  & (1202) &   -\frac{12x^3(1+3x+x^2)}{(1-x)^7}    \\ \hline \hline
\rule{0mm}{6mm} (1130) &   0   & (1121) &   \frac{x(12-14x+6x^2)}{(1-x)^3}   & (1112) &   \frac{4x(3+4x+x^2+x^3)}{(1-x)^5}    \\ \hline \hline
\rule{0mm}{6mm} (1103) &   \frac{20 x^3 (1+x)}{(1-x)^7}  & (1040) &   0 & (1031) &   0    \\ \hline \hline
\rule{0mm}{6mm} (1022) &   -\frac{2}{(1-x)^3}-\frac{4}{1-x}+\frac{1}{x}+6 & & & & \\ \hline \hline

\end{array}
\nonumber
\end{eqnarray}

\begin{eqnarray}
\begin{array}{|c|c||c|c||c|c|}
\hline 
\rule{0mm}{6mm} \vec{k}      & I_{\vec{k}}(x) & \vec{k} & I_{\vec{k}}(x) & \vec{k} & I_{\vec{k}}(x) \\ \hline \hline
\rule{0mm}{6mm} (1013) &   -\frac{2x (5-x+2x^2)}{(1-x)^5} & (1004) &   -\frac{10 x^3}{(1-x)^7}  &(0500) &   -\frac{x^6 (1+x+5x^2+5x^4+x^5+x^6)}{(1-x)^9 
(1+x+x^2+x^3+x^4)}   \\ \hline \hline
\rule{0mm}{6mm} (0410) &   \frac{x^4 (1+3x+x^2)}{(1-x)^7}    & (0401) &   \frac{x^5(1+6x+6x^2+x^3)}{(1-x)^9}  & (0320) &   -\frac{2x^3}{(1-x)^5}  \\ \hline
\hline
\rule{0mm}{6mm} (0311) &   -\frac{x^3 (1+9x+9x^2+x^3)}{(1-x)^7}  & (0302) &   -\frac{2x^5 (3+8x+3x^2)}{(1-x)^9}&  (0230)& 0 \\ \hline \hline
\rule{0mm}{6mm} (0221) &   \frac{3x^2(1+x)}{(1-x)^5}  &(0212) &   \frac{6x^3 (1+3x+x^2)}{(1-x)^7}  & (0203) & \frac{14x^5 (1+x)}{(1-x)^9}\\ \hline \hline
\rule{0mm}{6mm} (0140) &   0   & (0131) &   0    & (0122) &   -\frac{x(1+4x+x^2)}{(1-x)^5}  \\ \hline \hline
\rule{0mm}{6mm} (0113) &   -\frac{10x^3 (1+x)}{(1-x)^7} & & & &  \\ \hline \hline
\end{array}
\nonumber
\end{eqnarray}

\begin{eqnarray}
\begin{array}{|c|c|}
\hline 
\rule{0mm}{6mm} \vec{k}      & \sum_{m\neq 0}N_{\vec{k},m}x^{m} \\ \hline \hline

\rule{0mm}{6mm} (0104) &   -\frac{14 x^5}{(1-x)^9}    \\ \hline \hline
\rule{0mm}{6mm} (0050) &   0    \\ \hline \hline
\rule{0mm}{6mm} (0041) &   0    \\ \hline \hline
\rule{0mm}{6mm} (0032) &   0    \\ \hline \hline
\rule{0mm}{6mm} (0023) &   \frac{x(1+x)}{(1-x)^5}    \\ \hline \hline
\rule{0mm}{6mm} (0014) &   \frac{5x^3}{(1-x)^7}    \\ \hline \hline
\rule{0mm}{6mm} (0005) &   \frac{x^5(5+2x+2x^2+5x^3)}{(1-x)^9 (1+x+x^2+x^3+x^4)}    \\ \hline \hline

\end{array}
\nonumber
\end{eqnarray}

\section{Local ${\cal B}_{3}$}

\begin{eqnarray}
\begin{array}{|c|c||c|c||c|c|}
\hline 
\rule{0mm}{6mm} \vec{k}            & I_{\vec{k}}(x) & \vec{k} & I_{\vec{k}}(x)
& \vec{k} & I_{\vec{k}}(x) \\ \hline \hline
\rule{0mm}{6mm} (3000) & 0   & (2100) & \frac{1}{x^{2}}+\frac{2}{x}-\frac{3x}{1-x}    & (2010) &  0      \\ \hline \hline
\rule{0mm}{6mm} (2001) &-\frac{1}{x}+\frac{3x}{1-x}  & (1200) &\frac{1}{x}-\frac{x(6x^{2}-14x+10)}{(1-x)^3}    &(1110) &-\frac{1}{x^{2}}+\frac{4x}{1-x}         \\ \hline \hline
\rule{0mm}{6mm} (1101) &\frac{6x^3-14x^2+12x}{(1-x)^3}  & (1020) &0  & (1011) &\frac{1}{x}-\frac{2x}{1-x}         \\ \hline \hline
\rule{0mm}{6mm} (1002) & -\frac{2x}{(1-x)^{3}}   & (0300) &-\frac{x(1+x^{2})}{(1-x)^5(1+x+x^{2})}       & (0210) &-\frac{1}{x}+\frac{x(4x^{2}-9x+6)}{(1-x)^3}         \\ \hline \hline
\rule{0mm}{6mm} (0201) &\frac{x(1+x)}{(1-x)^5}   & (0120) &0& (0111) & -\frac{4x^3-9x^2+7x}{(1-x)^3}        \\ \hline \hline
\rule{0mm}{6mm} (0102) &-\frac{2x^2}{(1-x)^5}   & (0030) &   0& (0021) & 0        \\ \hline \hline
\rule{0mm}{6mm} (0012) & \frac{x}{(1-x)^3}  &(0003) & \frac{x^3(1+x)}{(1-x)^5(1+x+x^2)}   & &     \\ \hline \hline
\end{array}
\nonumber
\end{eqnarray}

\begin{eqnarray}
\begin{array}{|c|c||c|c|}
\hline 
\rule{0mm}{6mm} \vec{k}  & I_{\vec{k}}(x) & \vec{k} & I_{\vec{k}}(x)\\ \hline \hline
\rule{0mm}{6mm} (4000) &  0   & (3100) &  \frac{1}{x^3}+\frac{2}{x^2}+\frac{3}{x}-\frac{4x}{1-x} \\ \hline \hline
\rule{0mm}{6mm}  (3010) &  0   & (3001) &  -\frac{1+x-2x^2-4x^3}{(1-x)x^2}
\\ \hline \hline
\rule{0mm}{6mm}(2200) &  \frac{2+x(1+x)(6-26x-15x^2+61x^3-32x^4)}{(1-x)^3 x^2 (1+x)} & (2110) &  -\frac{1+x+x^2-3x^3-3x^4}{(1-x)x^3}    \\ \hline \hline
\rule{0mm}{6mm} (2101) &  -\frac{6}{x}+\frac{10x(6-9x+4x^2)}{(1-x)^3}& (2020) &  0    \\ \hline \hline
\rule{0mm}{6mm}(2011) &  -\frac{3x}{1-x}+\frac{1}{x^2}+\frac{2}{x} & (2002) &  -\frac{x(15-2x-11x^2+8x^3)}{(1-x)^3 (1+x)}  \\ \hline \hline
\rule{0mm}{6mm}(1300) &  \frac{1}{x}-\frac{2x(15-40x+48x^2-27x^3+6x^4)}{(1-x)^5}  & (1210) &  -\frac{3+3x-27x^2-7x^3+54x^4-30x^5}{(1-x)^3 x^2}   \\ \hline \hline
\rule{0mm}{6mm} (1201) &  \frac{2x(20-43x+50x^2-27x^3+6x^4)}{(1-x)^5}  & (1120) &  0   \\ \hline \hline
\rule{0mm}{6mm}(1111) &  36-\frac{8}{(1-x)^3}-\frac{28}{1-x}+\frac{7}{x}&  (1102) &  -\frac{2x(5-x+2x^2)}{(1-x)^5} \\ \hline \hline
\rule{0mm}{6mm}(1030) &  0  & (1021) &  0    \\ \hline \hline
\end{array}
\nonumber
\end{eqnarray}

\begin{eqnarray}
\begin{array}{|c|c||c|c|}
\hline 
\rule{0mm}{6mm} \vec{k}      & I_{\vec{k}}(x) & \vec{k} & I_{\vec{k}}(x) \\ \hline \hline
\rule{0mm}{6mm} (1012) &  \frac{2x(6-7x+3x^2)}{(1-x)^3}   & (1003) &  \frac{4x^2}{(1-x)^5}    \\ \hline \hline
\rule{0mm}{6mm} (0400) &  -\frac{x(1+2x+4x^2+2x^3+x^4)}{(1-x)^7 (1+x)^3}   & (0310) &  -\frac{1}{x}+\frac{x(20-57x+70x^2-40x^3+9x^4)}{(1-x)^5}    \\ \hline \hline
\rule{0mm}{6mm} (0301) &  \frac{x (1+3x+x^2)}{(1-x)^7}   & (0220) &  \frac{1-4x^2-3x^3+6x^4+3x^5-4x^6}{(1-x)^3 x^2 (1+x)}    \\ \hline \hline
\rule{0mm}{6mm} (0211) &  -\frac{x(26-62x+73x^2-40x^3+9x^4)}{(1-x)^5}   & (0202) &  -\frac{2x^2 (2+7x +12x^2+7x^3+2x^4)}{(1-x)^7 (1+x)^3}    \\ \hline \hline
\rule{0mm}{6mm} (0130) &  0    & (0121) &  -\frac{1}{x}+\frac{x(6-9x+4x^2)}{(1-x)^3}    \\ \hline \hline
\rule{0mm}{6mm} (0112) &  \frac{3x(2-x+x^2)}{(1-x)^5}  & (0103) &  \frac{5x^3}{(1-x)^7}    \\ \hline \hline
\rule{0mm}{6mm} (0040) &  0  & (0031) &  0    \\ \hline \hline
\rule{0mm}{6mm} (0022) &  -\frac{x}{(1-x)^3 (1+x)}   & (0013) &  -\frac{2x^2}{(1-x)^5}    \\ \hline \hline
\rule{0mm}{6mm} (0004) &  -\frac{2 x^4(1+3x+x^2)}{(1-x)^7 (1+x)^3}  &&   \\ \hline \hline
\end{array}
\nonumber
\end{eqnarray}

\begin{eqnarray}
\begin{array}{|c|c||c|c|}
\hline 
\rule{0mm}{6mm} \vec{k}      & I_{\vec{k}}(x) & \vec{k} & I_{\vec{k}}(x) \\ \hline \hline
\rule{0mm}{6mm} (5000) &   0  &(4100) &   \frac{1+x+x^2+x^3-4x^4-5x^5}{(1-x)x^4} \\ \hline \hline
\rule{0mm}{6mm} (4010) &   0   &(4001) &   -\frac{1+x+x^2-3x^3-5x^4}{(1-x)x^3}    \\ \hline \hline
\rule{0mm}{6mm} (3200) &   \frac{4+4x +9x^2-91x^3-21x^4+195x^5-110x^6}{(1-x)^3x^3}  &(3110) &   -\frac{1+x+x^2+x^3-4x^4-4x^5}{(1-x)x^4}    \\ \hline \hline
\rule{0mm}{6mm} (3101) &   -\frac{9+13x-93x^2-89x^3+290x^4-150x^5}{(1-x)^3 x^2}  & (3020) &   0    \\ \hline \hline
\rule{0mm}{6mm} (3011) &   \frac{1+x+x^2-3x^3-4x^4}{x^3-x^4}  & (3002) &   5(-\frac{2}{(1-x)^3} - \frac{6}{1-x}+\frac{1}{x}+8)   \\ \hline \hline
\rule{0mm}{6mm} (2300) &   \frac{4}{x^2}+\frac{30}{x}-\frac{x(300-948x+1217x^2-720x^3+165x^4)}{(1-x)^5} &(2210) &   -2\frac{3+2x+6x^2-60x^3+4x^4+100x^5-60x^6}{(1-x)^3 x^3}    \\ \hline \hline
\rule{0mm}{6mm} (2201) &   -\frac{20-100x-265x^2+1160x^3-1597x^4+965x^5-225x^6}{(1-x)^5 x}   &(2120) &   0    \\ \hline \hline
\rule{0mm}{6mm} (2111) &   2\frac{6+8x-60x^2-33x^3+149x^4-80x^5}{(1-x)^3 x^2}   & (2102) &   -\frac{x(177-415x+485x^2-265x^3+60x^4)}{(1-x)^5}   \\ \hline \hline
\rule{0mm}{6mm} (2030) &   0  &&  \\ \hline \hline

\end{array}
\nonumber
\end{eqnarray}

\begin{eqnarray}
\begin{array}{|c|c||c|c|}
\hline 
\rule{0mm}{6mm} \vec{k}      & I_{\vec{k}}(x) & \vec{k} & I_{\vec{k}}(x) \\ \hline \hline
\rule{0mm}{6mm} (2021) &   0   & (2012) &   -\frac{6}{x}+\frac{10x(6-9x+4x^2)}{(1-x)^3} \\ \hline \hline
\rule{0mm}{6mm} (2003) &   \frac{x(12-3x+5x^2)}{(1-x)^5}   & (1400) &   \frac{1}{x}-2\frac{x(35-133x+260x^2-287x^3+186x^4-66x^5+10x^6)}{(1-x)^7}    \\ \hline \hline
\rule{0mm}{6mm} (1310) &   -2\frac{3+5x-70x^2+15x^3+317x^4-556x^5+370x^6-90x^7}{(1-x)^5 x^2}   & (1301) &   2\frac{x(50-147x+281x^2-297x^3+189x^4-66x^5+10x^6)}{(1-x)^7}   \\ \hline \hline
\rule{0mm}{6mm} (1220) &   \frac{2+3x^2-29x^3+9x^4+37x^5-24x^6}{(1-x)^3 x^3}    & (1211) &   2\frac{13-65x-104x^2+575x^3-826x^4+509x^5-120x^6}{(1-x)^5 x}  \\ \hline \hline
\rule{0mm}{6mm} (1202) &   -\frac{6x(5+6x^2-2x^3+x^4)}{(1-x)^7}  &(1130) &   0    \\ \hline \hline
\rule{0mm}{6mm} (1121) &   -\frac{3+3x-27x^2-7x^3+54x^4-30x^5}{(1-x)^3 x^2}    & (1112) &   \frac{12x(14-34x+40x^2-22x^3+5x^4)}{(1-x)^5}    \\ \hline \hline
\rule{0mm}{6mm} (1103) &   \frac{4x^2(7+x+2x^2)}{(1-x)^7}  & (1040) &   0    \\ \hline \hline
\rule{0mm}{6mm} (1031) &   0  & (1022) &   -\frac{2}{(1-x)^3}-\frac{4}{1-x}+\frac{1}{x}+6  \\ \hline \hline

\end{array}
\nonumber
\end{eqnarray}\begin{eqnarray}
\begin{array}{|c|c|}
\hline 
\rule{0mm}{6mm} \vec{k}      & \sum_{m\neq 0}N_{\vec{k},m}x^{m} \\ \hline \hline

\rule{0mm}{6mm} (0104) &   -\frac{14 x^4}{(1-x)^9}    \\ \hline \hline
\rule{0mm}{6mm} (0050) &   0    \\ \hline \hline
\rule{0mm}{6mm} (0041) &   0    \\ \hline \hline
\rule{0mm}{6mm} (0032) &   0    \\ \hline \hline
\rule{0mm}{6mm} (0023) &   \frac{x(1+x)}{(1-x)^5}    \\ \hline \hline
\rule{0mm}{6mm} (0014) &   \frac{5x^3}{(1-x)^7}    \\ \hline \hline
\rule{0mm}{6mm} (0005) &   \frac{x^5(5+2x+2x^2+5x^3)}{(1-x)^9 (1+x+x^2+x^3+x^4)}    \\ \hline \hline

\end{array}
\nonumber
\end{eqnarray}

\begin{eqnarray}
\begin{array}{|c|c|}
\hline 
\rule{0mm}{6mm} \vec{k}      & I_{\vec{k}}(x)  \\ \hline \hline
\rule{0mm}{6mm} (1013) &   -\frac{2x (5-x+2x^2)}{(1-x)^5}  \\ \hline \hline
\rule{0mm}{6mm} (1004) &   -\frac{10 x^3}{(1-x)^7}    \\ \hline \hline
\rule{0mm}{6mm} (0500) &   -\frac{x (1+x+5x^2+5x^4+x^5+x^6)}{(1-x)^9 (1+x+x^2+x^3+x^4)} \\ \hline \hline
\rule{0mm}{6mm} (0410) &   -\frac{1-7x-29x^2+168x^3-368x^4+429x^5-287x^6+104x^7-16x^8}{(1-x)^7 x}    \\ \hline \hline
\rule{0mm}{6mm} (0401) &   \frac{x(1+6x+6x^2+x^3)}{(1-x)^9} \\ \hline \hline
\rule{0mm}{6mm} (0320) &   \frac{2}{x^2}+\frac{10}{x}-\frac{x(59-195x+257x^2-155x^3+36x^4)}{(1-x)^5}    \\ \hline \hline
\rule{0mm}{6mm} (0311) &   -\frac{x(70-230x+437x^2-467x^3+299x^4-105x^5+16x^6)}{(1-x)^7}\\ \hline \hline
\rule{0mm}{6mm} (0302) &   -\frac{2x^2 (3+8x+3x^2)}{(1-x)^9}    \\ \hline \hline
\rule{0mm}{6mm} (0230) &   0    \\ \hline \hline 
\rule{0mm}{6mm} (0221) &   -\frac{6}{x}+\frac{3x(28-86x+110x^2-65x^3+15x^4)}{(1-x)^5}    \\ \hline \hline
\rule{0mm}{6mm} (0212) &   \frac{x(20-11x+27x^2-11x^3+5x^4)}{(1-x)^7} \\ \hline \hline
\rule{0mm}{6mm} (0203) &   \frac{14x^3 (1+x)}{(1-x)^9}   \\ \hline \hline
\rule{0mm}{6mm} (0140) &   0   \\ \hline \hline
\rule{0mm}{6mm} (0131) &   0    \\ \hline \hline
\rule{0mm}{6mm} (0122) &   -\frac{x(26-62x+73x^2-40x^3+9x^4)}{(1-x)^5}\\ \hline
\hline
\rule{0mm}{6mm} (0113) &   -\frac{2x^2(8-x+3x^2)}{(1-x)^7}   \\ \hline \hline

\end{array}
\nonumber
\end{eqnarray}

\end{document}